\documentclass[twocolumn]{IEEEtran}

\usepackage{cite}

\ifCLASSINFOpdf
 
\else
 
\fi
\usepackage{graphicx}

\graphicspath{{images/}}
\usepackage{enumitem}
\usepackage{amsmath}
\usepackage{subfigure}
\usepackage{float}
\usepackage{cite}
\usepackage{svg}
\usepackage{amsmath}
\usepackage{graphics}
\usepackage{graphicx}
\usepackage{amssymb}
\usepackage{algorithm}

\usepackage{mathrsfs}
\usepackage{siunitx}
\usepackage{booktabs}
\usepackage{bm}
\usepackage{bbm}

\usepackage{float}
\usepackage{url}
\usepackage{amsfonts}
\usepackage[square,sort&compress,numbers]{natbib}
\usepackage{multirow}
\usepackage{textcomp}
\usepackage{supertabular}
\usepackage{array}
\usepackage{hyperref}

\newcommand{\tabincell}[2]{\begin{tabular}{@{}#1@{}}#2\end{tabular}}
\usepackage{enumerate}
\usepackage{graphicx}%
\usepackage{multirow}%
\usepackage{amsmath,amssymb,amsfonts}%
\usepackage{amsthm}%
\usepackage{mathrsfs}%
\usepackage{xcolor}%
\usepackage{textcomp}%
\usepackage{manyfoot}%
\usepackage{booktabs}%
\usepackage{algorithm}%
\usepackage{algorithmicx}%
\usepackage{algpseudocode}%
\usepackage{listings}%
\usepackage{hyperref}

\DeclareRobustCommand*{\IEEEauthorrefmark}[1]{%
    \raisebox{0pt}[0pt][0pt]{\textsuperscript{\footnotesize\ensuremath{#1}}}}

\begin{document}

\title{DeepPhysiNet: Bridging Deep Learning and Atmospheric Physics for Accurate and Continuous Weather Modeling}

\author{%
  \IEEEauthorblockN{%
    Wenyuan Li\IEEEauthorrefmark{1,3}\textsuperscript{$\dagger$},
    Zili Liu\IEEEauthorrefmark{1,2}\textsuperscript{$\dagger$},
    Keyan Chen\IEEEauthorrefmark{1},
    Hao Chen\IEEEauthorrefmark{1,2},
    Shunlin Liang\IEEEauthorrefmark{3},
    Zhengxia Zou\IEEEauthorrefmark{1}\textsuperscript{*} and 
    Zhenwei Shi\IEEEauthorrefmark{1}\textsuperscript{*}
  }%
  
  \IEEEauthorblockA{\IEEEauthorrefmark{1}Beihang University},~
  \IEEEauthorblockA{\IEEEauthorrefmark{2}Shanghai AI Laboratory}, ~
  \IEEEauthorblockA{\IEEEauthorrefmark{3}University of Hong Kong}
}

\maketitle
\begingroup\renewcommand\thefootnote{$\dagger$}
\footnotetext{Equal contribution}
\endgroup

\begin{abstract}
Accurate weather forecasting holds significant importance to human activities. Currently, there are two paradigms for weather forecasting: Numerical Weather Prediction (NWP) and Deep Learning-based Prediction (DLP). NWP utilizes atmospheric physics for weather modeling but suffers from poor data utilization and high computational costs, while DLP can learn weather patterns from vast amounts of data directly but struggles to incorporate physical laws.  Both paradigms possess their respective strengths and weaknesses, and are incompatible, because physical laws adopted in NWP describe the relationship between coordinates and meteorological variables, while DLP directly learns the relationships between meteorological variables without consideration of coordinates.
To address these problems, we introduce the \emph{DeepPhysiNet} framework, incorporating physical laws into deep learning models for accurate and continuous weather system modeling.  
First, we construct physics networks based on multilayer perceptrons (MLPs) for individual meteorological variable, such as temperature, pressure, and wind speed. Physics networks establish relationships between variables and coordinates by taking coordinates as input and producing variable values as output. The physical laws in the form of Partial Differential Equations (PDEs) can be incorporated as a part of loss function. Next, we construct hyper-networks based on deep learning methods to directly learn weather patterns from a large amount of meteorological data. The output of hyper-networks constitutes a part of the weights for the physics networks. Experimental results demonstrate that, upon successful integration of physical laws, DeepPhysiNet can accomplish multiple tasks simultaneously, not only enhancing forecast accuracy but also obtaining continuous spatiotemporal resolution results, which is unattainable by either the NWP or DLP.
DeepPhysiNet effectively combines the complementary strengths of NWP and DLP within a unified framework, facilitating seamless integration of the latest advancements from both paradigms and unlocking their full potential for weather forecasting. Code for DeepPhysiNet is available as an open source repository on GitHub \url{https://github.com/flyakon/DeepPhysiNet}.
\end{abstract}


\section{Introduction}\label{sec1}

Weather forecasting with various temporal and spatial scales has far-reaching effects on multiple aspects of human society. The accuracy of weather forecasting hinges on the accurate modeling of complex weather dynamics and systems. Meteorological experts have continuously devoted themselves to the development of weather models. The present stage of advancement has predominantly led to the emergence of two modeling paradigms. One encompasses the widely-recognized Numerical Weather Prediction (NWP) methods \cite{bauer2015quiet,lorenc1986analysis,stensrud2009parameterization}. 
The other encompasses data-driven machine learning, especially for Deep Learning-based Prediction (DLP) methods, which have notably gained momentum recently \cite{ben2023rise,ren2021deep,espeholt2022deep,ham2019deep,wu2023interpretable,bi2023accurate,ebert2023outlook,chen2023fengwu,kulichenko2023uncertainty,zou2023deep}. The success of these two paradigms can be ascribed to their powerful abilities in weather system modeling, along with other unique advantages. Nevertheless, each of these paradigms also presents its own set of challenges, demanding additional research to bridge the gap towards the ideal weather model.
 
The remarkable progress of NWP is closely linked to the breakthrough of classical physics and mathematics principles. The fundamental atmospheric equations, in the form of multiple Partial Differential Equations (PDEs), together with numerical optimization methods, form the cornerstone of modern NWP \cite{abbe1901physical}. This also endows NWP methods with a solid foundation in terms of physical interpretability. However, it is precisely due to the characteristics of modeling based on PDEs and numerical optimization methods that inherent limitations exist within NWP methods. These limitations include incomplete differential equations resulting from the highly nonlinear, chaotic nature of atmospheric systems,  the under-utilization of extensive historical observational data and discrete forecast results due to the intrinsic limitations of numerical optimization methods, as well as significant additional computational costs. Despite the aforementioned challenges, the development process of NWP has produced large-scale meteorological analysis and reanalysis datasets, such as ERA-5 reanalysis \cite{hersbach2020era5} and others, paving the way for the rapid advancement of data-driven methods.

\begin{figure*}
\centering
\includegraphics[width=\linewidth]{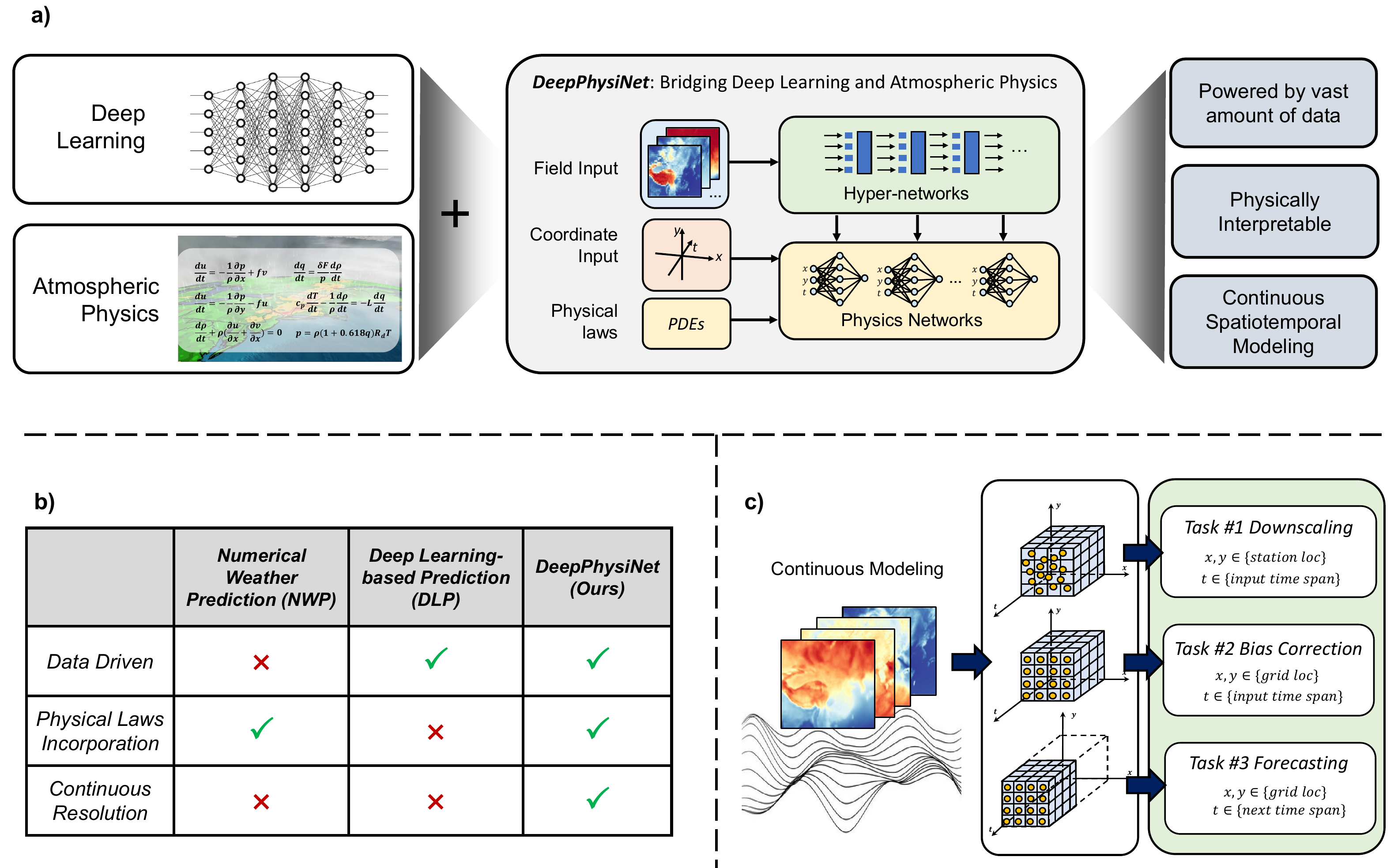}
\caption{ a) Overall of the proposed DeepPhysiNet, which incorporates atmospheric physics into deep learning methods for accurate and continuous weather modeling. It combines the advantages of NWP and DLP at the same time. b) 
Properties of DeepPhysiNet, compared to NWP and DLP. 
\textit{\textbf{Data Driven}.} It utilizes deep learning methods to extract spatiotemporal features from large-scale datasets.  \textit{\textbf{Physical Laws Incorporation}.} With the differentiable nature of neural networks, PDEs describing atmospheric physics can be incorporated into deep learning models with the form of loss function.  \textit{\textbf{Continuous Resolution.}} After trained, it can generate forecast results with a continuous resolution with the input of various sample points. c) Once trained, DeepPhysiNet is capable of performing various tasks, such as downscaling, bias correction, and forecasting, using multiple form of input coordinates.}
\label{fig:overview}
\end{figure*}

Thanks to the massive meteorological data generated by NWP and the observational data collected from various sensors, such as satellites and weather stations, the adoption of data-driven methods has emerged as a new opportunity. These methods, represented by Deep Learning-based Prediction (DLP), which has been widely demonstrated across numerous fields for knowledge extraction from large datasets \cite{lecun2015deep, ren2021deep, almalioglu2022deep, zhu2017deep, silver2017mastering, fawzi2022discovering, senior2020improved}, enable the direct utilization of vast data for weather forecasts. It offers not only simplified operations but also a significantly reduced demand for computational resources compared to NWP  \cite{espeholt2022deep, gronquist2021deep, lim2021time, gao2022earthformer, pathak2022fourcastnet}. In addition to their direct application in weather forecasting, deep learning methods can also enhance the accuracy or resolution of NWP results. This can be achieved by using deep learning for bias correction
\cite{hess2022physically, hess2022deep, han2021deep, karozis2023deep, gronquist2021deep, kim2021deep, mouatadid2023adaptive} or downscaling \cite{bano2020configuration, rampal2022high, harris2022generative, kumar2023modern}, thereby producing more refined forecasting results. However, the DLP also faces the limitation of producing forecasts at fixed resolutions due to the fixed training pipeline transplanted from computer vision tasks. Obtaining forecasts at different resolutions necessitates approximations through interpolation or re-training. Additionally, DLP can implicitly learn relationships between atmospheric variables from extensive data. Transferring the physical laws governing the atmosphere, in the forms of partial differential equations (PDEs) used in NWP, into deep learning methods poses significant challenges. As a result, DLP suffers not only from a lack of interpretability but also from the inability to guarantee full adherence to physical laws, potentially yielding results that defy conventional knowledge \cite{schultz2021can}. 

Recent advancements in physics-informed neural networks (PINNs) \cite{hao2022physics, raissi2019physics, raissi2020hidden,liu2022physics,kharazmi2021identifiability} present a successful attempt to incorporate physical laws into the data-driven methods. By employing Multilayer Perceptron (MLPs) to learn the mapping from spatiotemporal coordinates to state values and integrating soft constraints from differential equations, PINNs can offer continuous modeling of states described by PDEs through an optimization process, even with very limited observational data. As a result, optimization-based PINN methods have been extensively applied in various fields such as solving differential equations \cite{raissi2019physics} and modeling fluid dynamics \cite{raissi2020hidden}. There are also some early attempts in weather and climate modeling \cite{kashinath2021physics}.
Although neural networks can now be utilized to find numerical solutions for partial differential equations, this approach, much like numerical methods, fails to fully leverage the vast wealth of meteorological data \cite{raissi2019physics, liu2022novel, hao2022physics, giampaolo2022physics, kashinath2021physics, raissi2020hidden}.

To address these problems, we propose a unified framework, namely \emph{DeepPhysiNet}, which can incorporate atmospheric physics into deep learning methods for accurate and continuous weather modeling. In particular, as shown in Fig. \ref{fig:overview}a), it has two major components, physics networks and hyper-networks. The physics networks are constructed 
based on multilayer perceptrons (MLPs) for individual meteorological variables. The meteorological variables we focus include surface wind speed at 10 meters ($u,v$) in the east and north directions, air temperature at 2 meters ($T$), surface pressure ($p$), air density ($\rho$), and specific humidity ($q$). Physics networks establish relationships between variables and coordinates ($x,y,t$) by taking coordinates as input and producing variable values as output. Physics networks can leverage the differentiable nature of neural networks to construct a partial differential equation (PDE) loss based on atmospheric dynamics and thermodynamic equations.  Deep learning-based hyper-networks \cite{ha2016hypernetworks,chauhan2023brief} extracts spatiotemporal information from input meteorological field variables to generates the weights of physics networks. Under this framework, we transfer the weather information learned by the hyper-networks from a large amount of data to physical networks, thereby ensuring that physical networks obtain better weather modeling and physical embedding effects.

As shown in Fig. \ref{fig:overview}b), compared with  NWP and DLP, DeepPhysiNet possesses a large data capacity like deep learning methods, while also explicitly incorporating physical laws. This allows it to generate results that adhere more closely to the physical laws, akin to Numerical Weather Prediction (NWP) methods. More importantly, once the model is trained, it can produce continuous-resolution results by inputting coordinates in various arrangements, a benefit that is unattainable by either NWP or deep learning methods.

DeepPhysiNet can also be considered as a general modeling framework for weather systems. As depicted in Fig. \ref{fig:overview}c), it can be generalized to various downstream tasks by modifying the coordinate inputs of the physics networks. Therefore, we validate the effectiveness of the proposed framework through experiments on various tasks, including post-processing (downscaling and bias correction) and weather forecasting. The experimental results demonstrate that our p framework outperforms the currently operational weather forecasting systems on multiple tasks. 

\section{Framework}

\begin{figure*}[!htb]
\centering
\includegraphics[width=\linewidth]{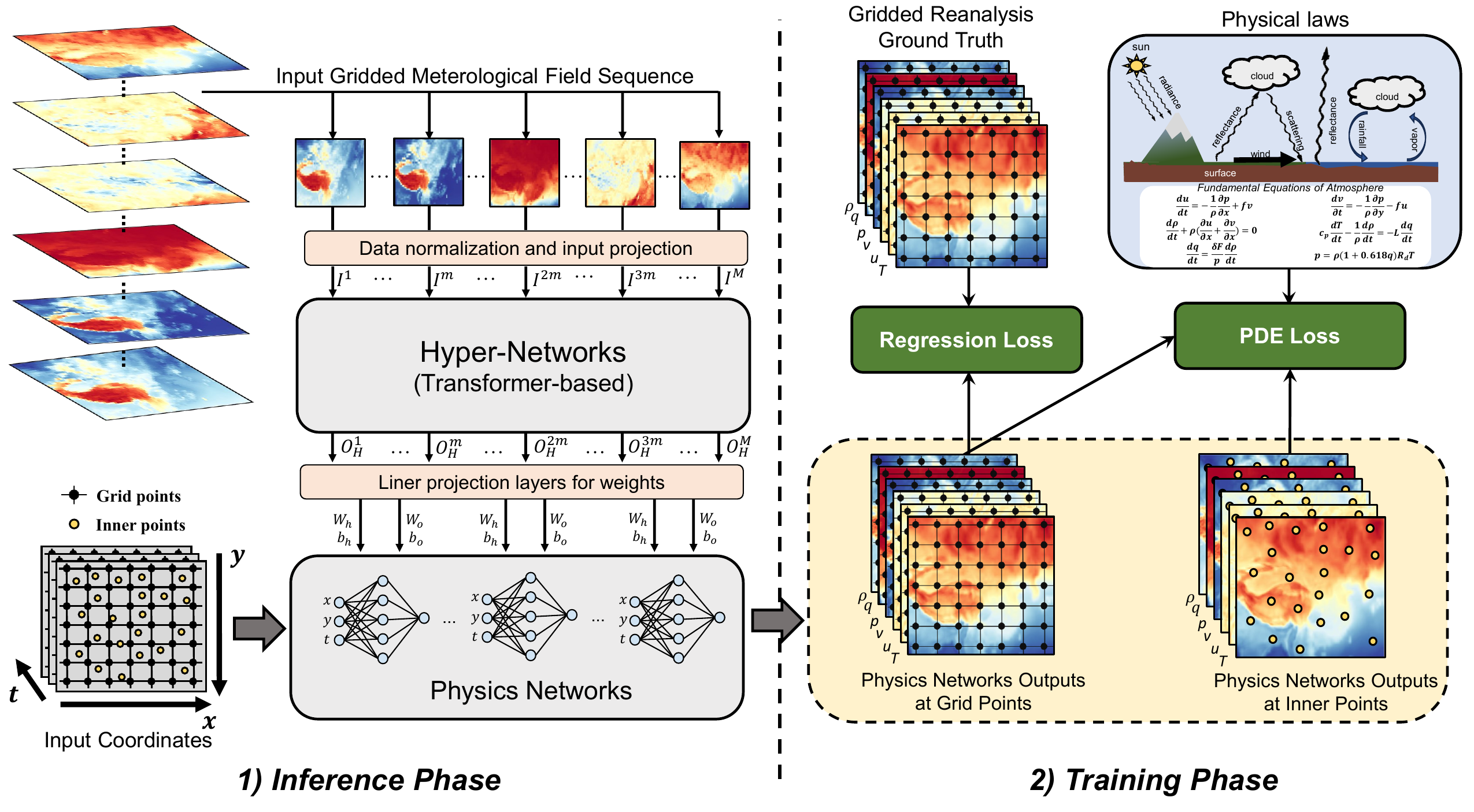}
\caption{Details of our proposed framework, DeepPhysiNet. It consists of two main components, hyper-networks and physics networks. The hyper-networks are responsible for extracting high-level spatiotemporal features from a large volume of historical meteorological field data and passing these features to the physics networks. The physics networks serve as neural solvers for partial differential equations describing near-surface meteorological variables. }
\label{fig:method}
\end{figure*}

In this section, we will introduce the general structure of the DeepPhysiNet framework, as shown in Fig. \ref{fig:method}.  

Both in the training and inference stages, the input of DeepPhysiNet comprises two parts:  meteorological field sequences and spatiotemporal coordinates. The former is fed into a Transformer-based \cite{vaswani2017attention, dosovitskiy2020image} hyper-networks and, through self-attention operations, generates high-level spatiotemporal features. The generated features are subsequently transformed into the shallow-layer weights of physics networks through a linear projection. Simultaneously, the spatiotemporal coordinates, comprising both grid points and inner points, pass through the multi-layer perceptrons (MLPs), which predict the desired variable states at corresponding input coordinates.   
Six surface meteorological variables are output by DeepPhysiNet, which include surface wind speed at 10 meters ($u,v$) in the east and north directions, air temperature at 2 meters ($T$), surface pressure ($p$), air density ($\rho$), and specific humidity ($q$). 
According to the vertical level, the input meteorological field data can be divided into two groups. $UU$, $VV$, $TT$, $Q$ and $z$ are all collected from various isobaric pressure levels, representing wind speed in the east and north directions, temperature, specific humidity and geopotential height, while $u$, $v$, $P$, $T$, $q$ and  $\rho$ are related the surface level, which are also the output variables of physics networks.

The physics networks, which can be treated as neural solvers for atmospheric partial differential equations, are guided by the received high-level spatiotemporal features and learn continuous mappings of coordinates and various target physical states. It also incorporates atmospheric physics in the form of soft constraints during the training phase. 
Specifically, the grid points refer to those which have corresponding ground truth, while inner points refer to those without ground truth sampled from arbitrary locations. For inner points,  we leverage the differentiable nature of neural networks to calculate partial differential equation (PDE) losses.  For grid points, besides the PDE losses, regression losses are also added. 
Due to the imposition of PDE constraints on inner points, we can change the input coordinates to the physics networks during the inference stage,  to obtain meteorological variables at any desired location.
It should be noted that for specific tasks, such as downscaling, bias correction, or forecasting, there are differences in the details of the model's input and output.

\section{Results}\label{sec2}
\subsection{Experiment settings}

The proposed DeepPhysiNet framework is capable of performing a wide range of meteorological tasks, relying on the input meteorological data and coordinates. We choose three specific tasks including downscaling, bias correction, and weather forecasting, to validate the effectiveness of the framework. These three tasks encompass two pivotal stages within operational forecasting: forecasting and post-processing (downscaling and bias correction). In the subsequent sections, we will offer a succinct overview of these three tasks and their corresponding experimental setups.

\textbf{Downscaling and Bias Correction:} 
The objectives of downscaling and bias correction are to post-process the forecast results generated by NWP. The downscaling task takes the NWP's coarse-resolution forecast results as input and generates fine-grained forecast results with higher resolutions or at station level. Meanwhile, the bias correction task aims to rectify the biases in the NWP's forecast results, yielding more accurate forecasts.

Both tasks can be performed using numerical methods by solving partial differential equations on a fine-grained grid, coupled with parameterizations. Similarly,  data-driven methods often draw inspiration from existing models and tasks in the computer vision domain, analogizing downscaling and bias correction to super-resolution \cite{chen2023continuous, saguy2023dblink} and denoising \cite{eom2023statistically, ghaddar2023denoising} tasks. However, both approaches are constrained by the inherent limitations of fixed super-resolution, and data-driven methods tend to disregard physical laws, resulting in poorer interpretability.

In contrast, DeepPhysiNet enables the downscaling of input coarse-resolution meteorological fields to a continuous high resolution while incorporating soft physical constraints. Specifically, we input forecast results with coarse spatiotemporal resolution into the hyper-networks and continuous spatiotemporal coordinates into the physics networks. High-resolution reanalysis data and physical soft constraints are utilized to supervise model training. It allows us to obtain corrected downscaled results at continuous resolutions.

\textbf{Weather Forecasting:} 
To further assess the weather system modeling capabilities of the proposed DeepPhysiNet, we conduct preliminary experiments to evaluate its forecasting ability for future weather states. Unlike operational numerical weather forecasts \cite{kalnay1998maturity} and recent deep learning-based forecast methods \cite{bi2023accurate, ebert2023outlook, chen2023fengwu}, our method does not primarily focus on comparing performance in weather forecasting. Instead, we aim to validate DeepPhysiNet's effectiveness in incorporating physical laws. We design experiments to predict meteorological data for the next 24 hours ($0h\to 24h$) based on discrete initial fields from the past 24 hours ($-24h\to 0h$). We employ discrete 24-hour reanalysis data for supervision, obtaining continuous spatiotemporal resolution forecasts. More importantly, we extrapolate from 24 hours to 48 hours by altering the input time coordinate $t$, where only physical laws are used as supervision.

Additionally, we have designed specific experiments to analyze the physical interpretability of DeepPhysiNet. Further details about these experiments can be found in Appendix \ref{app:experiments}.

\subsection{Data and Study Area}

In the validation of the DeepPhysiNet framework, we employ three types of datasets. Firstly, we utilize Numerical Weather Prediction (NWP) results obtained from the TIGGE project \cite{bougeault2010thorpex}, which is designed for inter-comparing various countries' or regions' integrated forecasting systems (IFS). In this paper, we employ integrated forecast results from the United States National Centers for Environmental Prediction (NCEP IFS) and the European Centre for Medium-Range Weather Forecasts (ECMWF IFS). Specifically, NCEP IFS serves as the input for DeepPhysiNet in the tasks of continuous downscaling and bias correction. ECMWF IFS, being one of the most effective numerical forecasting models to date, serves as a crucial benchmark method in our experiments. All NWP results from TIGGE exhibit a spatial resolution of 0.5 degrees, a temporal resolution of 6 hours, and a forecast period of 360 hours.

For the tasks of continuous downscaling and bias correction, the input data of DeepPhysiNet are from NCEP IFS. The forecast results are released at 00:00 and 12:00 each day, but we only use the data released at 00:00 as input. We separate the forecast data into distinct input groups with step of 24 hours. Each group's forecast time can be represented as $[24 n, 24 (n+1)]$, where n is selected from the range $\{0,1,...,15\}$. The input time forecast in hyper-networks and physics networks is $24n$. The meteorological variables input in the physics networks are obtained by interpolation of the input data according to coordinates and time. In addition to the input data, the input of hyper-networks also includes geographical information about the study area, such as its altitude, location, and proportion of land / sea. All input variables are re-sampled to the resolution of 1  degree.

Reanalysis data from ECMWF (ERA5) \cite{hersbach2020era5} is employed as both the labels for training the DeepPhysiNet framework and the input for weather forecasting task. The ERA5 dataset exhibits a spatial resolution of 0.25 degrees and a temporal resolution of 1 hour.

Lastly, we incorporate a subset of observational data to assess the performance of DeepPhysiNet. The observational data is derived from Weather2K dataset \cite{zhu2023weather2k}. From this dataset, we uniformly select 200 observational stations for validation purposes. The validation data covers the period from January 1, 2021, to August 1, 2021. 

\begin{figure}[!htb]
    \centering
    \includegraphics[width=0.75\linewidth]{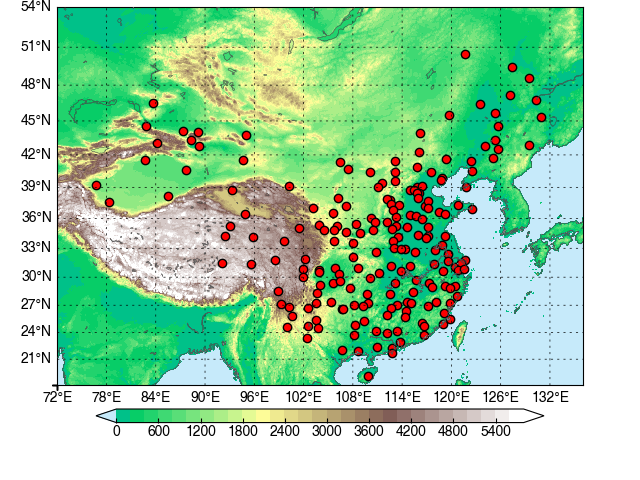}
    \caption{Study area with boundary with $72^{\circ} E$ to $136^{\circ} E$ and $18^{\circ} N$ to $54^{\circ} N$.  The red circles represent the observational station from Weather2K dataset.}
    \label{fig:study_area}
\end{figure}

Fig. \ref{fig:study_area} illustrates the locations of the observational stations and study area with boundary with $72^{\circ} E$ to $136^{\circ} E$ and $18^{\circ} N$ to $54^{\circ} N$ used in this experiment.  The red circles represent the observational station from Weather2K dataset.  More detailed information about the used datasets can be found in Appendix \ref{app:data}.

\subsection{Evaluate Metric}
We utilize the surface temperature (T), wind speed (SPD), and relative humidity (RH) as the variables for validation. These three variables can be calculated from the six meteorological variables from outputs of DeepPhysiNet.
The calculation formulas for wind speed and relative humidity are as follows:
\begin{equation}
    SPD=\sqrt{u^2+v^2},
\end{equation}
where $u,v$ represent the surface wind speed at 10 meters in the east and north directions, respectively.

\begin{equation}
    RH = \frac{q}{(1-q)w_s},
\end{equation}
where $w_s$ is the saturation mixing ratio, which can be calculated from: $r = 0.622 {e}/{p - e}$. $e$ represents saturation water vapor (partial) pressure. It can be calculated as follows:
\begin{equation}
    e=6.112 e^{\frac{17.67T'}{T' + 243.5}},
\end{equation}
where $T' = T-273.15$ is the temperature in degrees Celsius. 
Briefly, relative humidity is related to the three variables temperature, air pressure, and specific humidity that DeepPhysiNet outputs.

We employ root mean square error (RMSE) and the correlation coefficient (COR) as evaluation metrics.
The formula for calculating RMSE is as follows:
\begin{equation}  
 RMSE = \sqrt{\frac{1}{n}\sum_{i=1}^{n}(Y_{i} - X_{i})^{2}}.
\end{equation}
$n$ represents the number of samples. $Y_{i}$ denotes the label values. $X_{i}$ denotes the forecasting values.

The formula for calculating the correlation coefficient (COR) is as follows:
\begin{equation}
     COR = \frac{\sum_{i=1}^{n}(Y_{i} - \bar{Y})(X_{i} - \bar{X})}{\sqrt{\sum_{i=1}^{n}(Y_{i} - \bar{Y})^{2}}\sqrt{\sum_{i=1}^{n}(X_{i} - \bar{X)}^{2}}}.
\end{equation}
$\bar{Y}$ denotes the mean of label values, while
$\bar{X} $ denotes the mean of forecasting values.
A larger correlation coefficient value indicates a higher degree of agreement between label and output data.

In addition, we also use Kullback-Leibler divergence (KL-divergence, denoted as $D_{KL}$) to quantify the difference between two probability distributions. Here is the formula for KL-divergence:
\begin{equation}
    D_{KL}(P\|Q) = \sum_{i} P(i) \log\left(\frac{P(i)}{Q(i)}\right).
\end{equation}
In this formula, \(P\) and \(Q\) represent two probability distributions, while \(P(i)\) and \(Q(i)\) denote the probabilities of event \(i\) in these distributions.

\subsection{Downscaling with Continuous Resolutions}

In this section, we first evaluate the continuous downscaling performance of DeepPhysiNet using station observational data.  Despite DeepPhysiNet utilizing input data with a resolution of 1 degree and labels with a resolution of 0.25 degrees, it is capable of providing accurate results with continuous resolutions on station-level without re-training. It can be performed just by giving the coordinates of the observational stations to the physics networks during the inference stage. Table \ref{tab:stn_var_metric} displays station site-level downscaling results for wind speed (SPD), temperature (T), and relative humidity (RH) of various methods.

\begin{table}
    \centering
    \caption{Station site-level downscaling results for wind speed (SPD), temperature (T) and relative humidity (RH) of various methods.}
    % \resizebox{\linewidth}{!}{
    \begin{tabular}{c|cccccc}
		\toprule
		 \multirow{2}{*}{Method} & \multicolumn{2}{c}{SPD} & \multicolumn{2}{c}{T} & \multicolumn{2}{c}{RH} \\
         & RMSE & COR & RMSE & COR & RMSE & COR\\
		 \midrule
		 NCEP IFS & 2.055 & 0.439 & 4.178 & 0.937 & 19.364 & 0.711 \\
            ECMWF IFS  & 1.844 & 0.430 & 4.099 & 0.944 & 17.471 & 0.757\\
           DeepPhysiNet & \textbf{1.686} & \textbf{0.443} & \textbf{3.788} & \textbf{0.949}  & \textbf{16.686} & \textbf{0.761}  \\
		 \bottomrule
	\end{tabular}
 	% }
	\label{tab:stn_var_metric}
\end{table}

It is evident that our method, which utilizes NCEP's coarse-grained forecast results as input, substantially improves the accuracy and precision of weather forecasts, resulting in more refined station site-level forecasts. What's even more noteworthy is that our method, whether it pertains to input data or labels, relies solely on coarse-grained meteorological data and PDE constraints to ensure the validity of predictions at any location. Once the training is completed, there is no need for any special operations to achieve downscaling to the station-level.

\begin{figure*}
    \centering
    \includegraphics[width=0.9\linewidth]{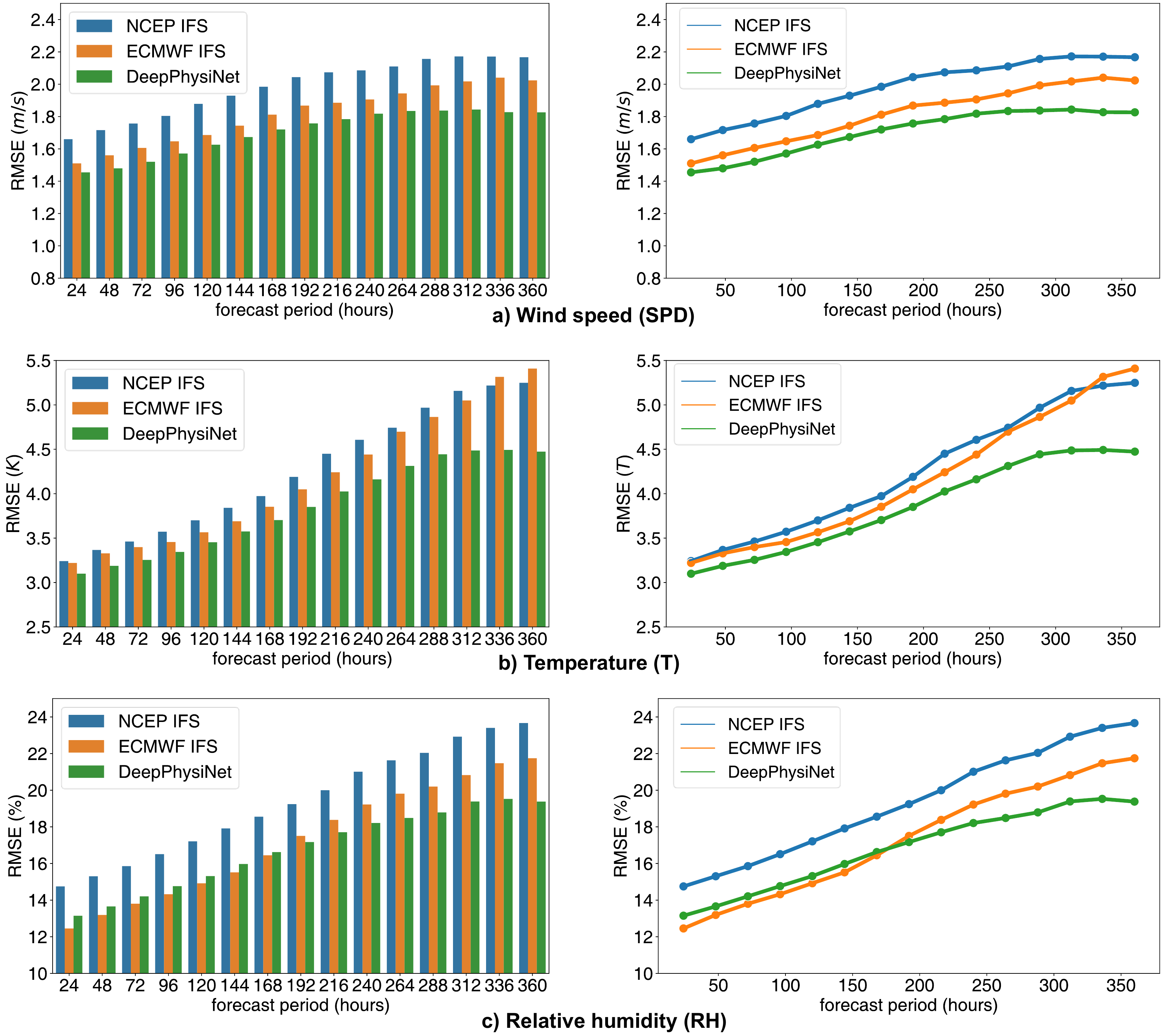}
    \caption{Station-level downscaling results for wind speed (SPD), temperature (T), and relative humidity (RH) at different forecast periods with 24-hour intervals.}
    \label{fig:stn_bar}
\end{figure*}

Fig. \ref{fig:stn_bar} employs bar and line charts to illustrate the station site-level downscaling results for wind speed, temperature, and relative humidity at different forecast duration with 24-hour intervals. Our method significantly reduces forecast errors and outperforms current state-of-the-art forecasting models. Notably, the improvement in wind speed forecasting is most significant. This is attributed to the fact that actual wind speeds exhibit the most substantial variability, as compared to temperature and relative humidity.
The experimental results also indicate that our method exhibits a more substantial improvement in forecasts beyond the 192-hour steps. This is due to the absence of cumulative errors in our method as opposed to numerical forecasting methods, which tend to have larger errors for longer forecast periods.

\begin{figure*}[!htb]
    \centering
    \includegraphics[width=0.95\linewidth]{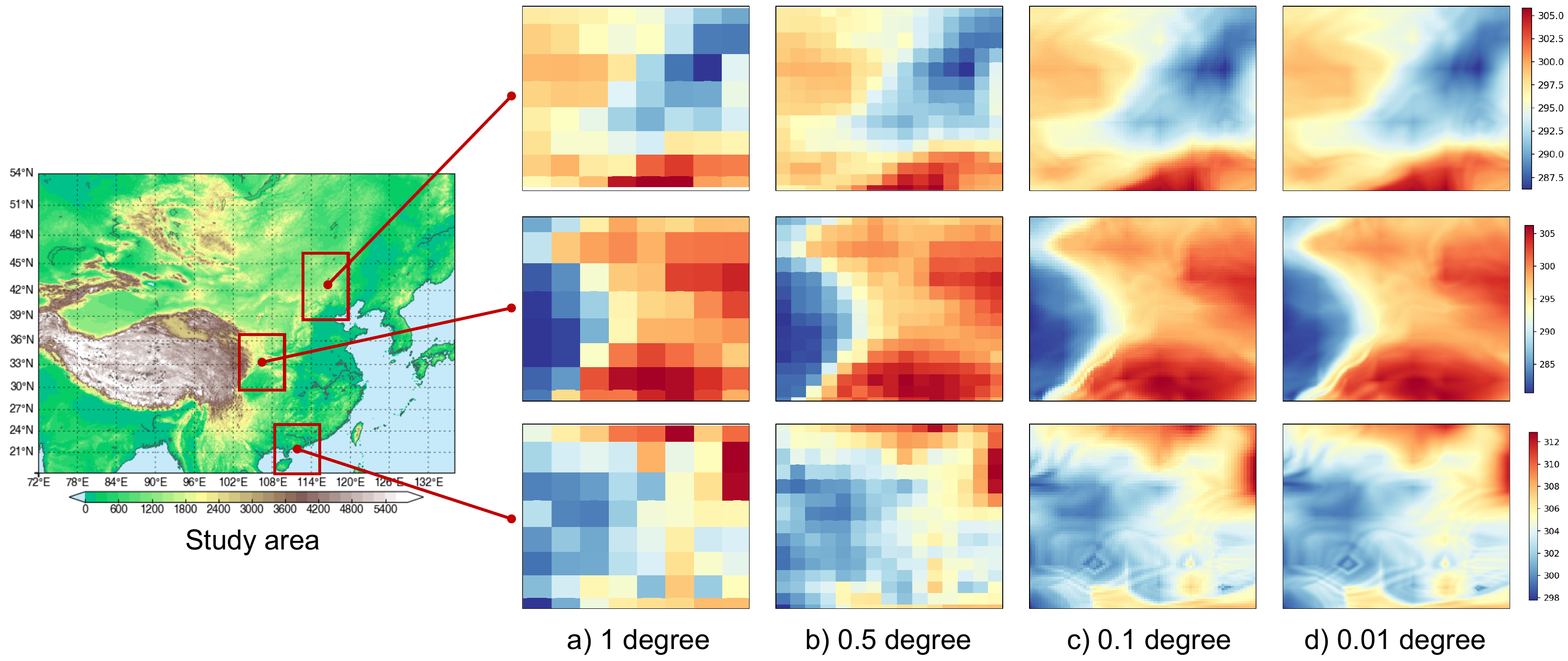}
    \caption{Continuous downscaling temperature forecasts for sub-regions at resolutions of 1 degree, 0.5 degrees, 0.1 degrees, and 0.01 degrees.}
    \label{fig:partial_vis}
\end{figure*}

Furthermore, we select three sub-regions with varying terrain characteristics in the study area. Using the DeepPhysiNet method, we obtain downscaling temperature forecasts for these sub-regions at resolutions of 1 degree, 0.5 degrees, 0.1 degrees, and 0.01 degrees, as shown in Fig. \ref{fig:partial_vis}. It can be observed that in the 1-degree and 0.5-degree outputs, pronounced jagged patterns are evident. However, as the resolution increases, our method effectively enhances the refinement of forecast results. What's more significant is that from the figure, it can be seen that in regions with only one result in low-resolution data, our method can generate diverse results, all of which conform to the original data's patterns. A more intuitive demonstration is that the obtained temperature field data corresponds to real-world conditions as altitude and terrain change.

\subsection{Bias Correction at Grid Points}
In this section, we assess the bias correction performance of DeepPhysiNet on grid points. We utilize forecast data from NCEP and ECMWF for 2021 to 2022 as a comparison. The spatial resolution of the test data is 1 degree with a temporal resolution of 6 hours (consistent with the input NCEP forecast data).

\begin{figure*}[!htb]
    \centering
    \includegraphics[width=0.9\linewidth]{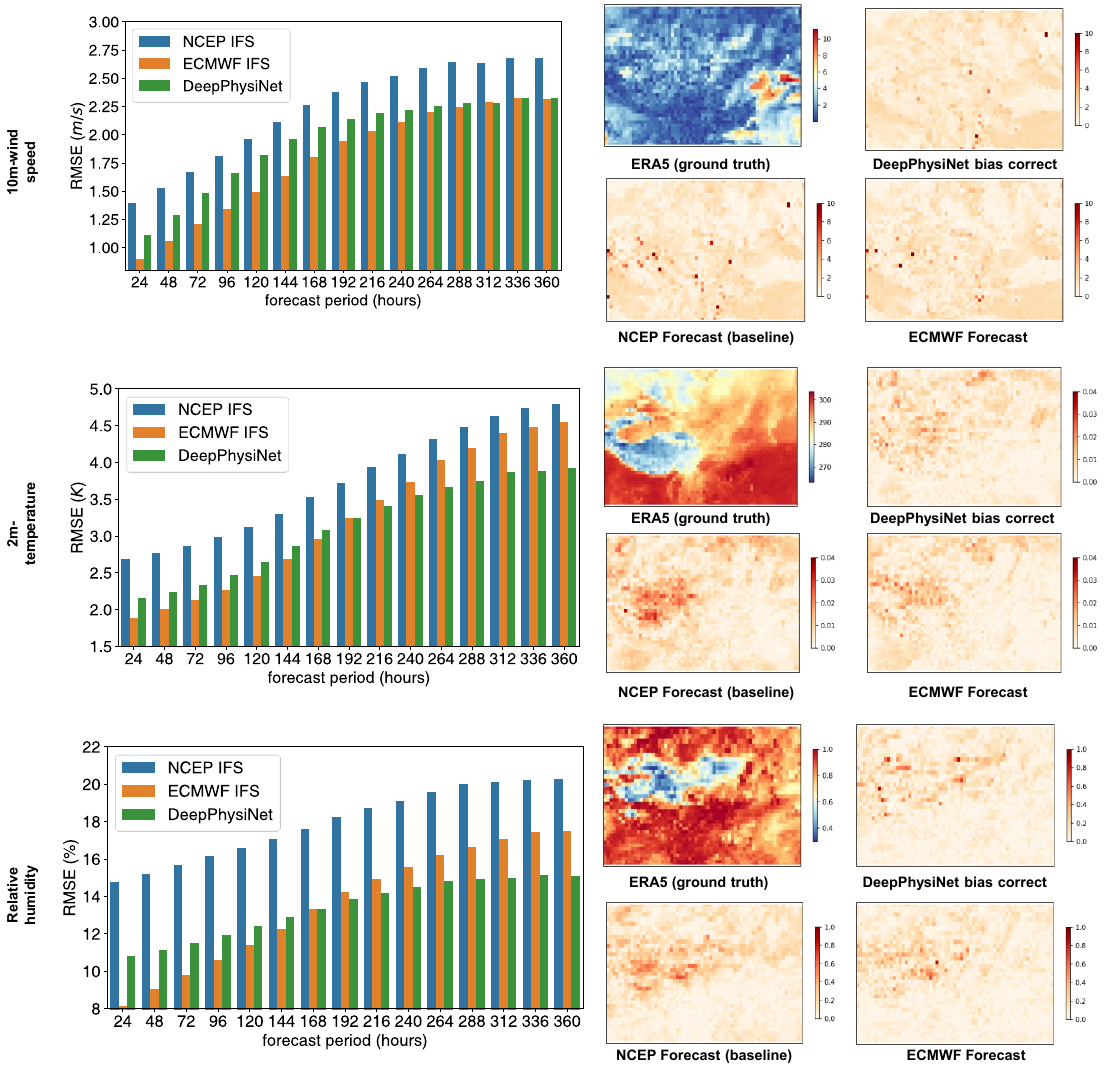}
    \caption{The bias correction performance at grid points for wind speed (SPD), temperature (T), and relative humidity (RH) at different forecast periods with 24-hour intervals. For each variable, some visualization results are given for comparison, including data from ERA5 as ground truth, NCEP IFS results as baseline, results corrected by our proposed  DeepPhysiNet and ECMWF IFS results. The results are obtained by calculating errors with ERA5 data (ground truth). }
    \label{fig:grid_bar}
\end{figure*}

Table \ref{tab:grid_var_metric} presents the bias correction results for different methods. During the inference process, we select two months from both 2021 and 2022 for each season as validation data. Additionally, NCEP forecasts serve as input data for our method, whereas ECMWF forecasts are not used for model training but rather as a benchmark for the best current forecasting model. Fig. \ref{fig:grid_bar} illustrates the bias correction performance for wind speed, temperature, and relative humidity at different forecast periods with intervals of 24 hours. 

\begin{table*}[!htb]
    \centering
    \caption{Bias correction performance at grid points for wind speed (SPD), temperature (T) and relative humidity (RH) of various methods.}
    \resizebox{\linewidth}{!}{
    \begin{tabular}{c|c|cccccccc}
		\toprule
		\multirow{2}{*}{Variables} & \multirow{2}{*}{Method} & \multicolumn{2}{c}{Spring} & \multicolumn{2}{c}{Summer} & \multicolumn{2}{c}{Fall} & \multicolumn{2}{c}{Winter} \\
        & & RMSE & COR & RMSE & COR & RMSE & COR & RMSE & COR\\
		 \midrule
		 \multirow{3}{*}{SPD} & NCEP IFS & 2.477 & 0.598 & 2.446 & 0.583 & 2.221 & 0.682  & 2.188 & 0.746\\
           & ECMWF IFS  & 1.916 & 0.696 & 1.854 & 0.695 &  1.876 & 0.742 & 1.843 & 0.802  \\
           & DeepPhysiNet &  2.056 & 0.659 & 2.073 & 0.629 &  2.004 & 0.718 & 1.965 & 0.791\\
           \midrule
          \multirow{3}{*}{T} & NCEP IFS & 3.683 & 0.960 &  3.301 & 0.927 & 3.663 & 0.972 & 4.221 & 0.967\\
           & ECMWF IFS & 3.244 & 0.969 & 2.753 & 0.950 & 3.330 & 0.977 & 3.876 & 0.972\\
           & DeepPhysiNet  & 3.139 & 0.970 & 2.646 & 0.952 & 3.288 & 0.978 &  3.577 & 0.976\\
           \midrule
           \multirow{3}{*}{RH} & NCEP IFS  & 19.152 & 0.750 & 16.592 & 0.793 & 16.583 & 0.748 & 16.583 & 0.748   \\
           & ECMWF IFS &  15.254 & 0.834 & 12.794 & 0.854 & 13.634 & 0.807 &  13.882 & 0.788\\
           & DeepPhysiNet  & 13.969 & 0.841 & 11.887 & 0.860 & 13.138 & 0.800 & 14.225 & 0.756\\
		 \bottomrule
	\end{tabular}
 	}
	\label{tab:grid_var_metric}
\end{table*}

Overall, our method exhibits noticeable bias correction performance on input NCEP forecasts. For temperature and relative humidity, our method still outperforms ECMWF forecasts, yielding the best results. However, for wind speed, there is still some gap compared to ECMWF forecasts. This is primarily because the wind speed in grid data is a result of substantial averaging, whereas our method, due to PDE constraints, tends to capture instantaneous wind speed even in the training process, which does not fully correspond to the ground truth from ERA5 data. Through comparisons of station data and grid data, our method yields superior performance on station data. This can be partly attributed to our ability to effectively integrate atmospheric physics into the forecasting process, rendering predictions more consistent with real-world patterns. Moreover, grid data represents averaged wind speeds and may not always conform to physical laws.

In addition, Fig. \ref{fig:grid_bar} also give some visualization results for each variable. The results are obtained by calculating errors with ERA5 data (ground truth). It can be seen from the results that after bias correction by our method, the forecast results are more closer with ERA5 than NCEP IFS and perform more consistently with ECMWF IFS.

\subsection{Weather Forecast}
Besides post-processing forecast results, our method can also enable direct weather forecasting. In this section, we utilize ERA5 data with a spatial resolution of 1 degree and a temporal resolution of 6 hours for weather forecasting, specifically predicting weather conditions for the next 48 hours. However, distinct supervision information is employed during training for forecasts within the first 24 hours and those between 24 and 48 hours. For forecasts within the initial 24 hours, we incorporated corresponding ERA5 data and physical laws as supervision. In contrast, for the 24-hour to 48-hour forecasts, only the PDE losses are added, making predictions beyond 24 hours a form of extrapolation. 

\begin{table}[!htb]
    \centering
    \caption{Forecast performance for wind speed, temperature and relative humidity with the metric of RMSE.}
    % \resizebox{\linewidth}{!}{
    \begin{tabular}{c|c|ccc}
		\toprule
		 Period & Method & SPD & T & RH \\
		 \midrule
		\multirow{2}{*}{0 - 24h} & NCEP IFS & 1.403  & 2.435  & 13.953  \\
         & DeepPhysiNet & 1.409 &2.078&11.555 \\
         \midrule
         \multirow{2}{*}{24 - 48h} &NCEP IFS & 1.539 & 2.544 & 14.464 \\
         & DeepPhysiNet (ext) & 1.954 &3.107 &  13.542\\
		 \bottomrule
	\end{tabular}
 	% }
	\label{tab:forecast_metric}
\end{table}

\begin{figure*}
    \centering
    \includegraphics[width=0.9\linewidth]{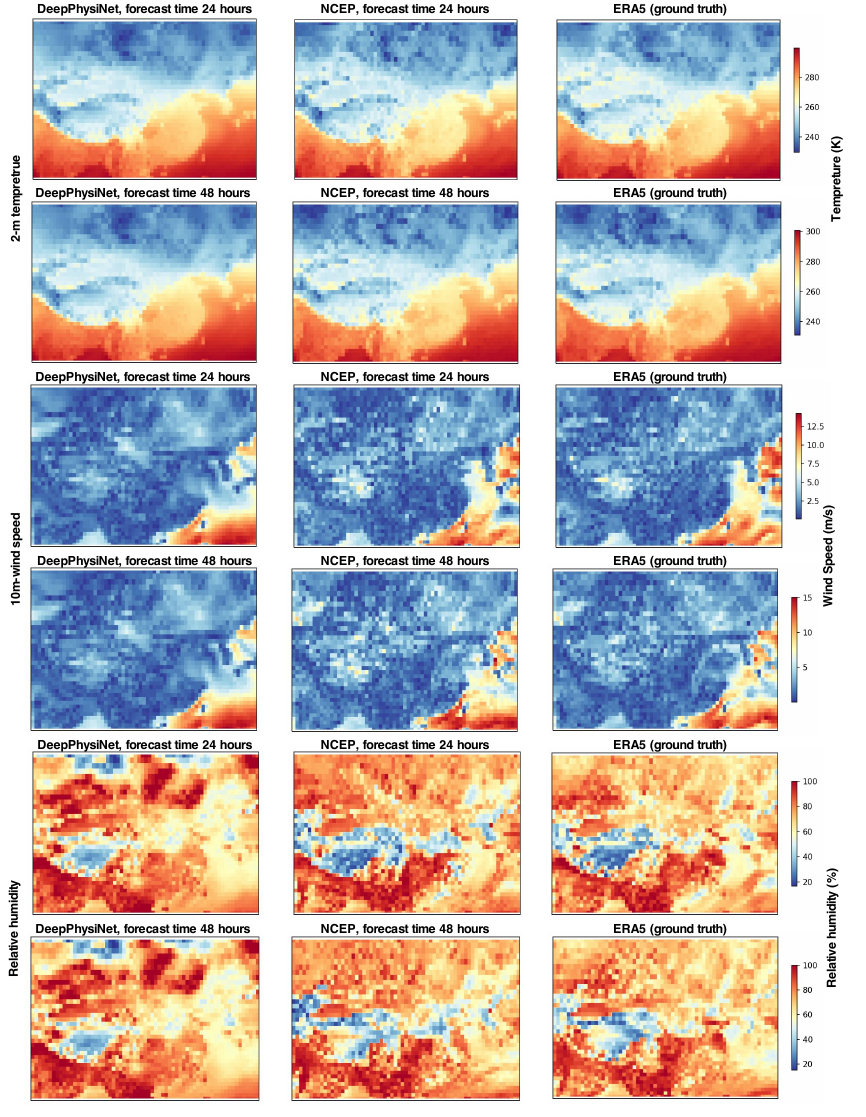}
    \caption{Visualization of forecast results. The 24 hours and 48 hours forecast results of 2m temperature and 10m wind speed from our proposed DeepPhysiNet, NCEP operational forecast and ERA5 ground truth are shown.}
    \label{fig:forecast_vis}
\end{figure*}

Table \ref{tab:forecast_metric} presents the results of weather forecasting, with "DeepPhysiNet (ext)" denoting the extrapolated forecasts for the 24-hour to 48-hour period, which rely solely on PDE loss for supervision. It is worth noting that even with this weak supervision, the performance of our method is not much worse than NCEP IFS, emphasizing its effectiveness. Moreover, for forecasts within the tightly constrained 24-hour range, our method significantly outperforms the NCEP IFS. Experimental results illustrate that our method is proficient at weather forecasting, and our proposed framework exhibits strong extrapolation capabilities. With the assistance of data and physical laws, we can employ straightforward extrapolation of sampled coordinates to obtain reasonable forecast results. 

Fig. \ref{fig:forecast_vis} illustrates the visualization of forecast results. From the forecast results, it can be observed that, firstly, in the 24-hour forecast, our method can produce results very close to ERA5 due to the constraint of strong supervision information. For forecasts from 24 to 48 hours, our method does not surpass NCEP IFS in terms of quantitative results. However, even under the soft constraint of only using PDEs losses, our method's forecast results are spatially very close to NCEP IFS and ERA5 data. This strongly demonstrates the effectiveness of incorporating PDEs. Since the PDEs we adopt are significantly simplified, the absence of any strong supervision information will inevitably cause larger deviations over time.

\subsection{Model Interpretability}

In this section, we analyze the roles played by atmospheric physics and input meteorological variables, validating the model's interpretability.

\begin{figure*}
    \centering
    \includegraphics[width=0.85\linewidth]{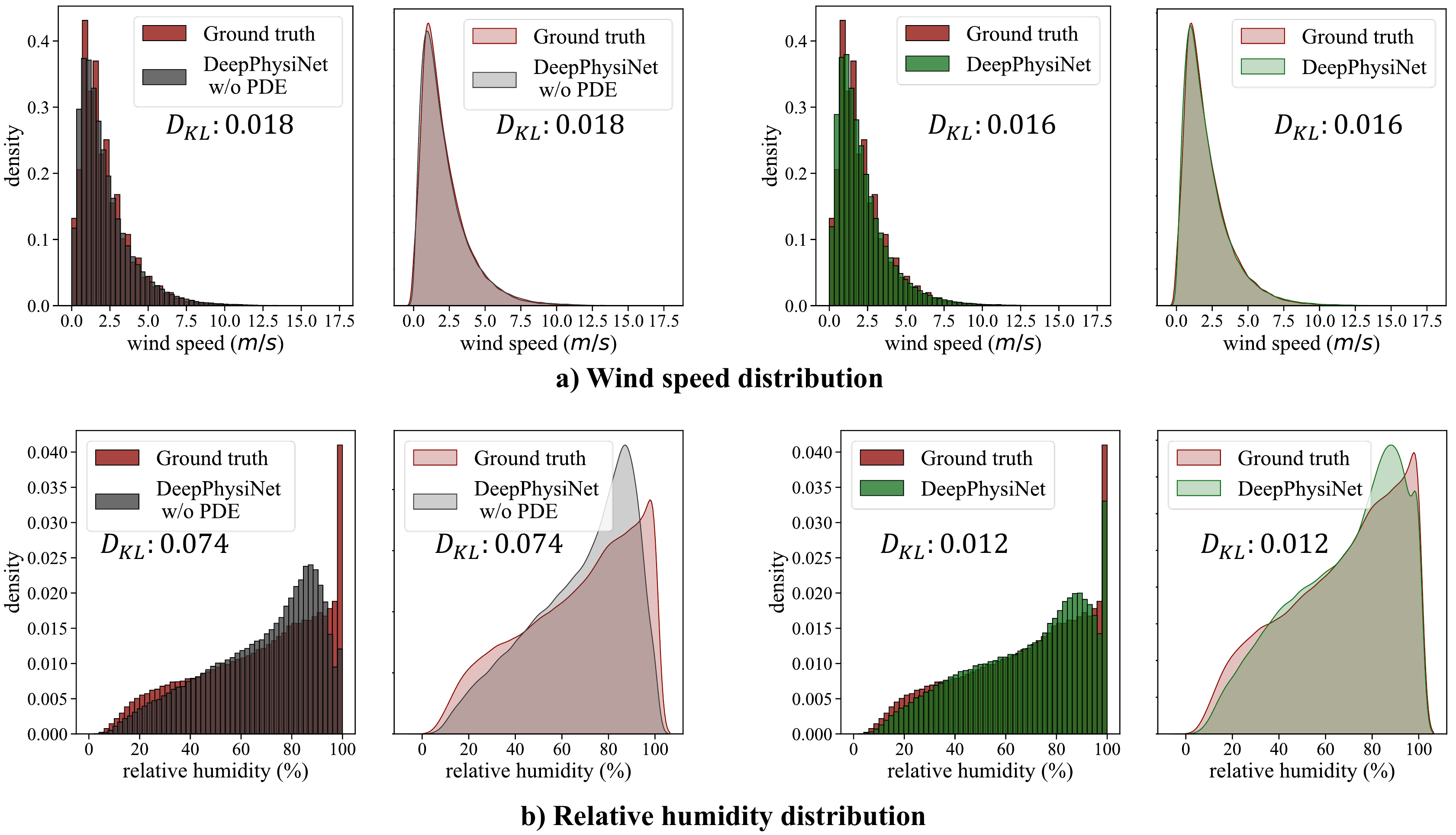}
    \caption{Comparison of the statistical distribution of downscaling results obtained with and without adding PDEs constraint.}
    \label{fig:density}
\end{figure*}

One essential role of physical laws is to ensure that forecast meteorological variables adhere to physical laws. Deep learning-based methods can only learn these patterns from extensive data. In this section, we compare the distribution of downscaled results from observational stations with ground truth to validate the influence of physical laws. We select wind speed and relative humidity as the validation variables, considering them as synthetic meteorological variables. Notably, relative humidity can simultaneously reflect the relationships among temperature, air pressure, specific humidity, and air density. To visually compare the impact of incorporating physical laws on the results, we also train a network model without including physical constraints (without PDE loss). We employ KL divergence as a metric to measure distribution distances. Fig. \ref{fig:density} illustrates the comparison of the statistical distribution of downscaling results obtained with and without adding PDEs constraint. 

It can be observed that the inclusion of physical constraints leads to results that are closer to the distribution of ground truth. Particularly for relative humidity, which involves multiple meteorological variables, incorporating physical constraints ensures that the variations of these variables align more closely with physical laws. For wind speed, incorporating physical constraints also yields a certain improvement. The incorporation of physical constraints provides an additional rational constraint to deep learning methods, ensuring more reasonable forecasts for various meteorological variables.  Certainly, it can attain a high degree of consistency even in the absence of physical laws (PDEs). This can be attributed to two main reasons. First, the deep learning method allows for the identification and assimilation of underlying physical patterns from the training data. Second, we have utilized one of the simplest form of atmospheric physics. Due to the chaotic nature of atmospheric motion, some processes are difficult to describe using explicit partial differential equations (PDEs). Consequently, the PDEs we employ may introduce considerable errors. For the sake of feasibility in model construction, we have further simplified equations, making our use of PDEs a rough constraint rather than a fully accurate representation.

We also employ the smoothGrad method \cite{smilkov2017smoothgrad} to investigate the contributions of different input variables to the model results. The contribution factor of each input variable $I$ can be expressed as follows:
\begin{equation}
    f=\frac{1}{N} \sum_{i=1}^{N} \nabla f(x + \epsilon_i)\,
\end{equation}
\(N\) is the number of samples, \(\epsilon_i\) represents noise samples drawn from a standard normal distribution, and \(\nabla f(x + \epsilon_i)\) indicates the computed gradient after adding noise \(\epsilon_i\). Among them, the larger the contribution factor $f$ obtained, the greater the impact of the input variable on the result.

\begin{figure*}
    \centering
    \includegraphics[width=0.85\linewidth]{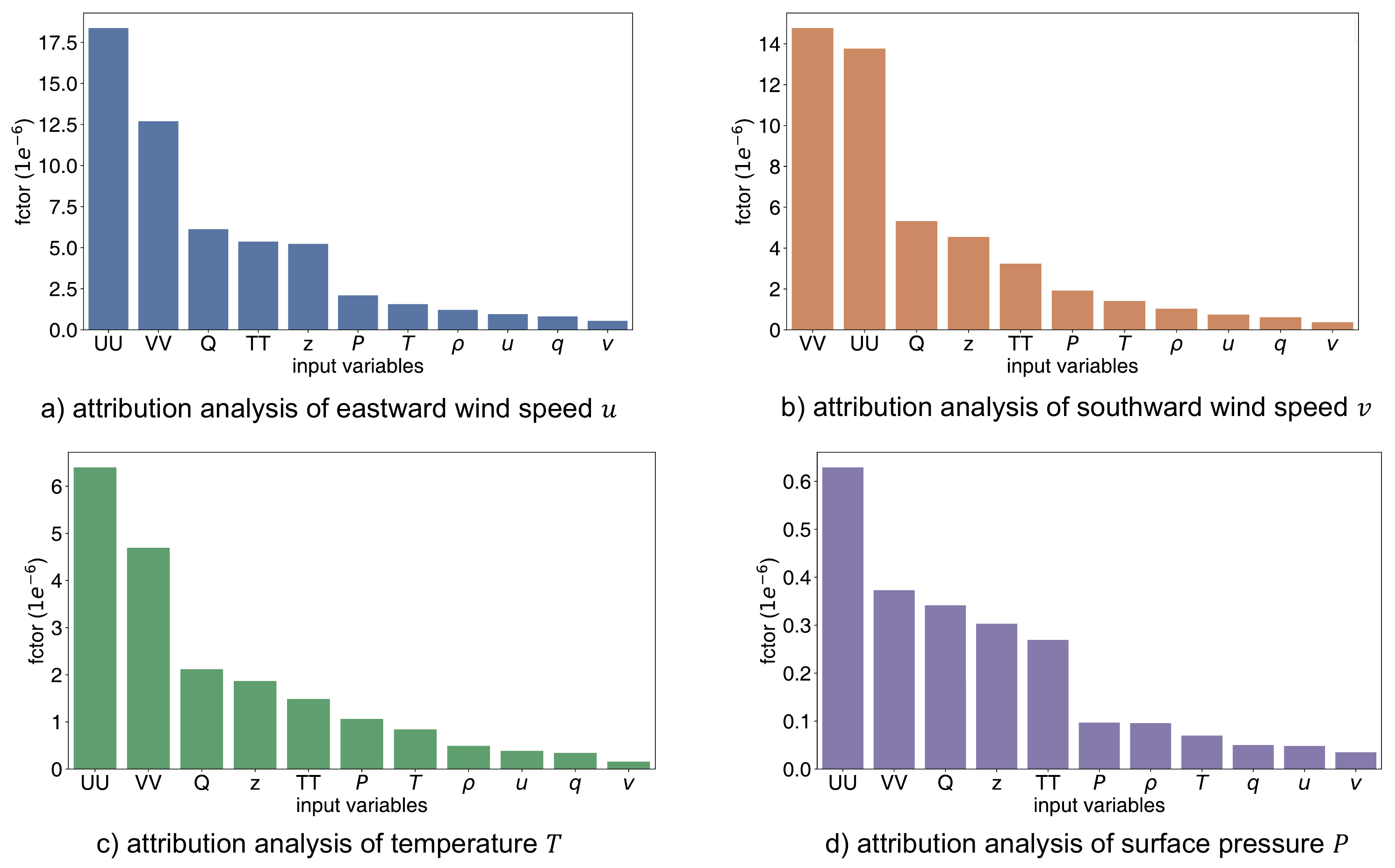}
    \caption{Contribution analysis results for eastward and southward  wind speed ($u$ and $v$), temperature ($T$), and surface pressure ($P$).}
    \label{fig:attr}
\end{figure*}

Fig \ref{fig:attr} illustrates the contribution factors of input variables for eastward wind speed $u$, southward wind speed ($v$), temperature ($T$), and surface pressure ($P$) individually. It can be observed that, for these four variables, the impact of pressure variables on the results is the most significant, while the influence of surface variables is comparatively small. Among the variables at different pressure levels, wind speed has the most substantial impact on the results. This is because, even though we primarily focus on surface variables, the surface weather conditions are closely linked to variations in the vertical atmosphere. The uneven heating of the atmosphere leads to the occurrence of various weather phenomena and induces horizontal and vertical motions at different atmospheric levels. Therefore, through the movement of the vertical atmosphere and changes in variables such as temperature, we can indirectly infer changes in surface variables.

\section{Method}
In this section, we present an in-depth description of the proposed framework, DeepPhysiNet, which includes the hyper-networks, physics networks, and the design of the training objective that incorporates atmospheric physics into deep learning methods.  

\subsection{Hyper-networks}

We first construct hyper-networks $H$ based on the transformer model \cite{vaswani2017attention,dosovitskiy2020image}, which uses low-resolution meteorological fields to extract spatial and time series features. The outputs of hyper-networks are used to calculate the physics network's weights. 
 
The low-resolution meteorological field data are reorganized into a format that can be inputted into the hyper-networks. Each pressure level is used as an independent input variable when multiple pressure levels are present in the input data. All variables are then transformed into vector form. Following this processing, there are $k$ input variables in each time step. For $m$ time steps, $k \times m$ input variables are available, denoted by $M_D$. In order to indicate which parts of hyper-networks' outputs will be used to calculate the weight of physics networks, a group of learnable parameters $W_L$ with the number of $M_L$ are connected to the end of input data.  After that processing, the input data can be recorded as $D\in R^{M\times N}$, where $N$ represents its dimension and $M=M_D+M_L$. 

In order for hyper-networks to recognize the times and relative positions relationship of the input variables, it is also necessary to encode the relative positions of the variables into the input data using positional encoding. The following formula is used to compute positional encoding:
\begin{equation}\label{eq:pe}
   \gamma(i)=\{\sin(2^0\pi i), cos(2^0\pi i), \ldots, sin(2^{N-1}\pi i), cos(2^{N-1}\pi i)\},
\end{equation}
where $i$ represents the position of each variables, with range of $(1,M)$.

The transformer consists of a series of identically-structured blocks. Each block contains a multi-head self-attention (MSA) module, a norm module and a feed-forward module. The self-attention module is the core part of the transformer network. The input of the block is the sum of the output of the previous block and the positional encoding. For the first block, it is the sum of the input data and position encoding, denoted as $I=PE+D$.

The self-attention module initially employs three learnable parameters $W^Q, W^K,W^V$ to map the input to a triplet $Q, K, V$:
\begin{equation}
    \begin{split}
        & Q=I\times W^Q,\\
        & K=I\times W^K,\\
        & V=I\times W^V.\\
    \end{split}
\end{equation}
The output of the self-attention module can be obtained by the following formula:
\begin{equation}
    Z=softmax(\frac{Q \times K^T}{\sqrt{N}}) \times V.
\end{equation}

We construct $n$ self-attention modules and separately calculate the results of $n$ outputs: $[Z_1,Z_2,\ldots,Z_{n-1},Z_n]$. A single fully connected layer is used to obtain the output of the multi-head self-attention (MSA) module:
\begin{equation}
    Z=[Z_1,Z_2,\ldots,Z_{n-1},Z_n] \times W^O.
\end{equation}

In order to make the training of hyper-networks more stable, we use layernorm \cite{vaswani2017attention}  to normalize the outputs of MSA. Finally, through the feed-forward network composed of multi-layer perceptrons (MLPs) and layer norm layer, we can get the output of the block.  After the concatenation of multiple layers of blocks, we can finally get the outputs of hyper-networks, denoted as $O_H$. 

\subsection{Physics networks}

The outputs of the hyper-networks can be used to calculate the weights of physics networks, which can apply the features by the hyper-networks to the physics networks and aid in establishing the mapping of coordinates and variables. Each output meteorological variable corresponds to a physics network, and each physics network has two sets of weights that are determined using hyper-networks.  In order to make the learning of physics networks more efficient, in addition to the input coordinates, we also use the interpolation method to obtain the reference values of the output variables from the low-resolution input meteorological field data according to the coordinates as an additional input to the physics network. In addition, time information is also added to the input of the network through position encoding.

For the input branch of coordinates, all network weights and biases are obtained from the hyper-network's outputs. Each physics network has two sets of weights and biases that need to be calculated, which are the weights from the input layer to the hidden layer: $W_h\in R^{D_I\times D_H}$, the biases from the input layer to the hidden layer: $b_h\in R^{D_H}$, the weights from the hidden layer to the output layer: $W_O\in R^{D_H\times D_O}$ and the bias from the hidden layer to the output layer: $b_o\in R^{D_O}$.

The dimension of the input coordinates is increased according to the Eq. \ref{eq:pe}, which is recorded as $C_I\in R^{D_I}$. Then, we calculate the output of this branch at specific coordinates with the following formula:
\begin{equation}
    \begin{split}
        & O_H=C_I\times W_H +b_H,\\
        & E_C=O_H \times W_O + b_O.
    \end{split}
\end{equation}

The input branch of time and the reference data share the same network architecture. First, the position encoding method is used to increase the dimension of the input data, which are recorded as $T_I$ and $V_I$ respectively, and then the linear projection layer transforms it:
\begin{equation}
    \begin{split}
        & E_T=T_I \times W_T + b_T,\\
        & E_V=V_I \times W_V + b_V.
    \end{split}
\end{equation}
$E_T$ and $E_V$ represent the embedding of input time and reference data, respectively. $W_T$ and $W_V$ represent weights of linear projection layer. $b_T$ and $b_V$ represent biases of linear projection layer.

The embedding obtained by the three input branches contains different information, but their dimensions are identical, so they can be combined to produce the input embedding as $I=E_C+E_T+E_V$. In order to accelerate network training, a residual MLP (ResMLP) \cite{touvron2022resmlp} module is incorporated to further transform the connected emending $I$. We employ two residual MLP modules in series to get the improved outputs for each variable. The detailed architecture of both hyper-networks and physics networks can be found in Appendix \ref{app:model_detail}.

\subsection{Training object}
The DeepPhysiNet's training object consists of two types of loss function: the regression loss and the partial differential equation (PDE) loss. For input coordinate points with corresponding ground-truth values, the regression loss between the physics network's outputs and the ground-truth can be calculated. 

Both variables on these two types of points these points must conform to the physical laws. By employing the partial differential equation (PDE) loss function, we impose constraints on the meteorological variables generated by the physics networks to adhere to the fundamental principles of physics. The partial differential equations (PDEs) chosen for this paper encompass the motion equation, the continuity equation, the energy equation, the water vapor equation, and the ideal gas state equation.

\subsubsection{Regression Loss}
The regression loss is defined based on smooth L1 loss \cite{girshick2015fast}, with the following specific form:
\begin{equation}
    S(A,\Tilde{A})=
        \begin{cases}
        0.5 (A - \Tilde{A})^2 / \beta, & \text{if } |A - \Tilde{A}| < \beta \\
        |A - \Tilde{A}| - 0.5 * \beta, & \text{otherwise}
        \end{cases},
\end{equation}
where $\beta$ is set to 0.1.

Assuming that $A$ represents any of the output variables $u, v, p, T, q, \rho$ and that $\Tilde{A}$ represents the ground truth, the regression loss function can be expressed as follows:
\begin{equation}
    L_r=\sum_{A\in \{u,v,p,T,q,\rho\}} \alpha_A* S(A,\Tilde{A}),
\end{equation}
where $\alpha$ is the balance coefficient. 
The regression loss function is used to constrain the output of the network at grid points.

\subsubsection{PDE Loss}
The PDE loss is calculated according to the basic equations of the atmosphere, including the motion equation, the continuous equation, the energy equation, the water vapor equation and the ideal gas state equation. In order to calculate PDE loss, it is necessary to convert the network's outputs to the original dimension of the variable. Nonetheless, the range gap between different variables may be relatively large, resulting in a significant disparity between PDE losses. To balance each PDE loss, we set the balance coefficients so that the value of the loss function is approximately the same. 

In this paper, the specific form of the motion equation we adopt is as follows:
\begin{equation}\label{eq:motion}
    \begin{split}
        & \frac{\mathrm{d} u}{\mathrm{d} t} = -\frac{1}{\rho} \frac{\partial p}{\partial x} +fv,\\
        & \frac{\mathrm{d} v}{\mathrm{d} t} = -\frac{1}{\rho} \frac{\partial p}{\partial y} - fu,
    \end{split}
\end{equation}
where $x,y,t$ represent the horizontal position and time, respectively.  $f$ represents the Coriolis coefficient, $f=2\Omega\sin\phi$. $\Omega$ denotes the earth's rotational angular velocity, which is equal to $7.29 \times 10^{-5}  s^{-1}$. $\frac{\mathrm{d}}{\mathrm{d} t}$ refers to $\frac{\partial }{\partial t} +u\frac{\partial }{\partial x} +v\frac{\partial }{\partial y}$.

The PDE loss of motion equation is defined as follows:
\begin{equation}
    \begin{split}
        & L_u =\alpha_u* MSE(\frac{\mathrm{d} u}{\mathrm{d} t},-\frac{1}{\rho} \frac{\partial p}{\partial x} + fv) ,\\
        & L_v=\alpha_v* MSE(\frac{\mathrm{d} v}{\mathrm{d} t},-\frac{1}{\rho} \frac{\partial p}{\partial y} - fu).
    \end{split}
\end{equation}
$MSE(\dots)$ represents the mean square error (MSE) loss function. $\alpha_u$ and $\alpha_v$ represent the balance factors, which are both set to $1e^3$.

The form of the continuous equation is as follows:
\begin{equation}
    \frac{\mathrm{d} \rho}{\mathrm{d} t}+\rho(\frac{\partial u}{\partial x}+ \frac{\partial v}{ \partial y})=0.
\end{equation}
The PDE loss of the continuous equation is defined as follows:
\begin{equation}
    \begin{split}
        L_c =\alpha_c* MSE(\frac{\mathrm{d} \rho}{\mathrm{d} t}+\rho(\frac{\partial u}{\partial x}+ \frac{\partial v}{ \partial y}),0).
    \end{split}
\end{equation}
$\alpha_c$ represents the balance factors, which is set to $1e^10$.

We simplify the thermodynamic equation by omitting external work on the atmosphere and considering only the heat released by water vapor condensation. The form of this equation is as follows:
\begin{equation}
    c_p\frac{\mathrm{d} T}{\mathrm{d} t}-\frac{1}{\rho}\frac{\mathrm{d} p}{\mathrm{d} t}=-L\frac{\mathrm{d} q}{\mathrm{d} t}.
\end{equation}
The PDE loss of the thermodynamic equation is defined as follows:
\begin{equation}
    L_e =\alpha_e* MSE(c_p\frac{\mathrm{d} T}{\mathrm{d} t}-\frac{1}{\rho}\frac{\mathrm{d} p}{\mathrm{d} t},-L\frac{\mathrm{d} q}{\mathrm{d} t}).
\end{equation}
$\alpha_c$ represents the balance factor, which is set to $1e^{1}$. 

The following is the form of the water vapor equation:
\begin{equation}
    \frac{\mathrm{d} q}{\mathrm{d} t}=\frac{\delta F}{p}\frac{\mathrm{d} p}{\mathrm{d} t}.
\end{equation}
\begin{equation}
\delta=
\begin{cases}
0,\  \frac{\mathrm{d} p}{\mathrm{d} t}< 0\ and\ q \ge q_s\\
1,\ else
\end{cases}
\end{equation}
\begin{equation}
    F=q_sT\frac{LR-c_pR_vT}{c_pR_vT^2+L^2q_s}.
\end{equation}
$q_s$ is the value for saturated specific humidity, which is determined using the saturated water vapor pressure and the empirical formula. The improved Stellen formula can be used to calculate the saturated vapor pressure $e_s$:
\begin{equation}
    e_s=6.112*\exp{\frac{17.67T'}{T'+243.5}},
\end{equation}
where $T' = T-273.15$ is the temperature in degrees Celsius. The saturated specific humidity can be calculated using the following empirical formula after obtaining the saturated water vapor pressure:
\begin{equation}
    q_s=\frac{0.622e_s}{p-0.378e_s}.
\end{equation}
The PDE loss of the water vapor equation is defined as follows:
\begin{equation}
    L_w =\alpha_w* MSE(\frac{\mathrm{d} q}{\mathrm{d} t},\frac{\delta F}{p}\frac{\mathrm{d} p}{\mathrm{d} t}).
\end{equation}
$\alpha_w$ represents the balance factor, which is set to $1e^{14}$. 

The form of the ideal gas state equation is as follows:
\begin{equation}
    p=\rho(1+0.608q) R_d T.
\end{equation}
The PDE loss of this equation is defined as follows:
\begin{equation}
    L_g =\alpha_g* MSE(p,\rho(1+0.608q) R_d T.
\end{equation}
$\alpha_g$ represents the balance factor, which is set to $1e^{-7}$. 
The total PDE loss is defined as $L_p= \alpha_u L_u + \alpha_v L_v + \alpha_c L_c + \alpha_e L_e + \alpha_w L_w$.

The PDE loss function can be calculated at grid points and interior points in order to constrain the network output to comply with physical laws. These points' PDE loss functions are denoted by the symbols $L_p^g$, and $L_p^i$, respectively.

The total loss function for network training is the sum of all the above loss functions:
\begin{equation}
    L=L_r^g+L_p^g+L_p^i.
\end{equation}
For each training step, the number of grid points and inner points we randomly selected are 20480 and 4096, respectively.

{\small
\bibliographystyle{IEEEtran}
\bibliography{refbib}
}

\bigskip

\newpage
\begin{onecolumn}
\begin{appendices}

\section{Experiments details for specific tasks}\label{app:experiments}

\textbf{Downscaling}. 
During the training phase, the hyper-networks accept coarse-grained meteorological field data as input. The field data, acquired from TIGGE, are generated by NCEP IFS. Table \ref{tab:tigge_var} shows detailed information about the input variables. According to the vertical level, the input data can be divided into two groups. $UU$, $VV$, $TT$, $Q$ and $z$ are all collected from various isobaric pressure levels, while $u$, $v$, $P$, $T$, $q$ and  $\rho$ are related the surface weather, which are also the output variables of physics networks.  

\begin{table*}[!htb]
    \centering
    \caption{Variables of TIGGE control forecast data.}
    \resizebox{\linewidth}{!}{
    \begin{tabular}{c|c|c|c|c|c|c|c}
		\toprule
		Variable & Description & Units & Level & Variable & Description & Units & Level \\
		\midrule
		$UU$ & U-component of wind & $m\cdot s^{-1}$ & \tabincell{c}{$1000 hpa, 925 hPa,$\\$ 850 hPa, 700 hPa, 500 hPa$} & $VV$ & V-component of wind & $m\cdot s^{-1}$ & \tabincell{c}{$1000 hpa, 925 hPa,$\\$ 850 hPa, 700 hPa, 500 hPa$} \\
		$TT$ & Temperature & $K$ & \tabincell{c}{$1000 hpa, 925 hPa,$\\$ 850 hPa, 700 hPa, 500 hPa$} & $Q$ & Specific humidity & $kg \cdot kg^{-1}$ & \tabincell{c}{$1000 hpa, 925 hPa,$\\$ 850 hPa, 700 hPa, 500 hPa$} \\
		$z$ & Geopotential height & $m$ & \tabincell{c}{$1000 hpa, 925 hPa,$\\$ 850 hPa, 700 hPa, 500 hPa$} &  \\
		\midrule
		
		$u$ & U-component wind at 10 meter & $m\cdot s^{-1}$ & surface & $v$ & V-component wind at 10 meter & $m\cdot s^{-1}$ & surface \\
		$P$ & Surface pressure & $Pa$ & surface & $T$ & Temperature at 2 meter & $K$ & surface \\
		$q$ & Specific humidity at 2 meter & $kg \cdot kg^{-1}$ & surface & $\rho$ & air density & $kg\cdot m^{-3}$ & surface \\
		\bottomrule
	\end{tabular}
 	}
	\label{tab:tigge_var}
\end{table*}

During the training phase, the variables from different isobaric pressure levels are regarded as different tokens of hyper-networks. Considering the fact that the ranges of different variables may vary greatly, We count the means and variances of different variables from the training data, and uniformly transform them into normal distributions and input them into the network to avoid biasing the training of the network due to different values of variables. However, when calculating the PDE losses, the normalized variables should be re-transformed as variables in PDEs need to use SI units. 

As for the input coordinates of physics networks, the horizontal coordinates $(x_{min}, y_{min})$ are used to represent the region's upper left corner and $(x_{max}, y_{max})$ are used to represent the region's lower right corner. For time coordinates $t$, the start time is denoted by $t_{min}$ and the end time is denoted by $t_{max}$. Usually, the $(x_{min}, y_{min})$ are set to $0$.

During training, coarse-grained input field data can be treated as discrete sampling values, with coordinates expressed as $x_s, y_s, p_s, t_s$. They can be represented by the following:
\begin{equation}\label{eq:input_coord}
    \begin{split}
        & \mathbf{x_s}=\{x_s \vert x_s=x_{min}+i*\Delta x \And x_s \le x_{max},i \in \mathbf{N}\},\\
        &  \mathbf{y_s}=\{y_s \vert y_s=y_{min}+i*\Delta y \And y_s \le y_{max},i \in \mathbf{N}\},\\
        &  \mathbf{t_s}=\{t_s \vert t_s=t_{min}+i*\Delta t \And t_s \le t_{max},i \in \mathbf{N}\},
    \end{split}
\end{equation}
where $\Delta x$, $\Delta y$ and $\Delta t$ represent resolutions of horizontal position and time. As we use ``meter'' and ``second'' as the units of horizontal position and time, $\Delta x$ and $\Delta y$ are set to $0.25 \times 10800=27000$. $\Delta t$ is set to $60*60*1=3600$. Because the ground-truth, ERA5 dataset, has a spatial resolution of $0.25^\circ$ and temporal resolution of $1 h$.

During the inference stage of downscaling tasks, however, the $\Delta x$, $\Delta y$ and $Delta t$ are not discrete anymore. We can adjust the $\Delta x$, $\Delta y$ and $Delta t$ to get continuous input coordinates corresponding to the station location in the weather2K dataset. These continuous coordinates then can be fed into physics networks to achieve downscaling at the station level.

\textbf{Bias correction}. 
For the bias correction task, the input data and coordinates are identical to those of downscaling tasks. The differences lie in the inference phase. During the inference phase of the bias correction task, we can just keep the $\Delta x$, $\Delta y$ and $Delta t$ as the same as input field data. 

\textbf{Forecasting}. 
Unlike the objectives of downscaling and bias correction, weather forecasting tasks involve predicting the corresponding meteorological variables at future time steps by inputting historical weather field data. Therefore, we leveraged the conventional setting of previous weather forecasting works and utilized ERA5 reanalysis data as input and supervision for our model. We utilized the subset of surface variables as in the downscaling task, where the historical values of six surface variables as inputs to hyper-networks for forecasting the corresponding variables at future time steps. Without loss of generality, during the training phase, we utilized a sequence of five frames of historical data with a temporal resolution of 6 hours as the input. We used the subsequent four frames of data, also with a temporal resolution of 6 hours, as the supervision for training. It is worth noting that the emphasis of the forecasting task lies in the ability to predict future time steps rather than improving spatial resolution. Therefore, we straightforwardly utilized $1^{\circ}$ spatial resolution data as both the input and output for the model.

In addition, we employed a specially designed input for the physical network based on the classic persistence method, tailored to the forecasting task. Specifically, the input for the physical network includes not only the spatiotemporal coordinates of future time steps (as described in Eq. \ref{eq:input_coord}) but also the persistence variable, denoted as $\mathbf{u}$, which captures the persistence of weather conditions. Therefore, for the forecasting task, the input-output representation of the model can be expressed as follows:
\begin{equation}
    \mathcal{F}^{out}_{t_0 : t_0 + 24h} = DeepPhysiNet(\theta, \omega; \mathcal{F}^{in}_{t_0 - 24h : t_0 : 6h}, \{\mathbf{x_s}, \mathbf{y_s}, \mathbf{t_s}, \mathbf{u_{t_0-24h : t_0}}\})
\end{equation}
where, $\mathcal{F}^{out}_{t_0 : t_0 + 24h}$ and $\mathcal{F}^{in}_{t_0 - 24h :6h: t_0}$ are output and input weather fields. $\theta$ and $\omega$ are learnable parameters of hyper-network and physics networks. $\mathbf{u_{t_0: t_0 + 24h}}$ is the interpolate persistent state vector of 6 surface variables that correspond to the input time span $[t_0-24h: t_0]$ and the spatiotemporal location $\{\mathbf{x_s}, \mathbf{y_s}\}$ of the output. By employing such inputs, we provide valuable prior information to the physical network, ensuring rapid convergence of the model. 

Besides, for extrapolation predictions beyond the supervised time window, we only need to adjust the time coordinate $\mathbf{t_s}$ in the input of the physical network to the corresponding extrapolation time point. The persistence factor remains the same as the input moment. This approach helps alleviate the issue of exponential error divergence caused by cumulative errors in the model's predictions. By maintaining the persistence factor unchanged, the model can retain the short-term variations and patterns in the forecasted variables, allowing it to capture and reproduce the high-frequency details of the predicted weather conditions over an extended period.

\section{Data Source}\label{app:data}
To validate the effectiveness of the DeepPhysiNet framework we proposed, experiments are conducted to distinguish among downscaling, bias correction, and forecasting tasks. Detailed descriptions of the data sources and preprocessing methods used for each task will be provided in this section. Table \ref{tab:dataoverall} displays the overall information of data used in this paper.

\begin{table*}
    \centering
    \caption{Table of Data Sources in this Paper.}
    \resizebox{\linewidth}{!}{
    \begin{tabular}{c|c|c|c|c|c}
		\toprule
		Data name & Data description & Time Span & Resolution & forecast period  & Task (Usage: Train (T) / Valid (V) \\
		\midrule
		TIGGE Archive  Control Forecast & Forecasts from NWP methods & 2008 $\to$ 2022& $1^{\circ}$ / 6h & \tabincell{c}{360h with \\
  a step of 6h} &\tabincell{c}{ Downscaling (T\&V) \\ Bias Correction (T\&V) \\ Forecast (V)} \\
        \midrule
            ERA-5 Reanalysis & Reanalysis & 2008 $\to$ 2022 & $0.25^{\circ}$ / 6h & - & \tabincell{c}{ Downscaling  (T)  \\ Bias Correction  (T\&V) \\ Forecast  (T\&V)}  \\
            \midrule
            Weather-2K & In-situ Observation & 2017 $\to$ 2021 & - / 1h & - & Downscaling (V) \\
            \midrule
            Geographic Auxiliary Data & Geographic Information & 2008 $\to$ 2022 & $1^{\circ}$ / 6h & - & \tabincell{c}{Downscaling (T\&V) \\ Bias Correction (T\&V) \\ Forecast (T\&V)} \\
		\bottomrule
	\end{tabular}
 	}
	\label{tab:dataoverall}
\end{table*}

\textbf{TIGGE Archive Control Forecast Data}. 
TIGGE is a crucial element of THORPEX, which is part of the World Weather Research Program. The primary objective of THORPEX is to accelerate advancements in the accuracy of high-impact weather forecasts spanning 1 day to 2 weeks, with a focus on benefiting humanity. TIGGE facilitates this goal by routinely generating global ensemble forecasts that extend up to approximately 14 days. These forecasts are produced by various meteorological centers worldwide, including ECMWF, JMA (Japan), Met Office (UK), CMA (China), NCEP (USA), MSC (Canada), Météo-France, BOM (Australia), CPTEC (Brazil), and KMA (Korea). The data generated from these centers is archived for research and operational purposes.

In this paper, the control forecast results at the surface and pressure levels produced by ECMWF and NCEP have been chosen as the model forecast outputs used for training and testing different task models. ECMWF's forecast outputs are widely regarded as state-of-the-art, while NCEP's forecast outputs are considered to have relatively lower accuracy. Specifically, we have downloaded forecast results that are initialized at 00:00 UTC each day, spanning a forecast time of 360 hours, with a spatial resolution of 1 degree, covering the geographical region of $72^{\circ} E$ to $136^{\circ} E$ and $18^{\circ} N$ to $54^{\circ} N$. 

\textbf{ERA-5 Reanalysis Data}. 
ERA5 reanalysis data is a weather dataset that combines past observations with models to generate consistent time series of multiple climate variables. It provides a comprehensive description of the observed climate as it has evolved during recent decades, on 3D grids at sub-daily intervals. ERA5 is the latest climate reanalysis produced by ECMWF, providing hourly data on many atmospheric, land-surface and sea-state parameters together with estimates of uncertainty. ERA5 data are available in the Climate Data Store on regular latitude-longitude grids at 0.25o x 0.25o resolution, with atmospheric parameters on 37 pressure levels. ERA5 has been available since 1940 and continues to be extended forward in time, with daily updates being made available 5 days behind real-time. 

\textbf{Weather2K}. 
The Weather2K dataset provides observational variables of 2,130 ground weather stations from January 2017 to August 2021 with a temporal resolution of 1 hour. For each station, 20 meteorological variables and 3 constants for position information are recorded. We use this dataset with the range of January 2021 to August 2021 to validate DeepPhysiNet's downscaling performance. Among the recorded variables, three variables, temperature, wind speed and relative humidity,  are selected for validation. 

\textbf{Geographic Auxiliary Data}. 
Geographic auxiliary data are utilized as the extra input of hyper-networks, which aims to help networks to identify the longitude, latitude and terrain of the study area. These data include: 1) longitude / latitude information of the study area extracted from ERA5 directly. 2) elevation map calculated using geopotential from ERA5. 3) land / sea mask data extracted from ERA5 directly.

\section{Model details}\label{app:model_detail}

\textbf{Hyper-networks}. 
The hyper-networks mainly consist of a series of multi-head self-attention (MSA)  with identically-structure. However, for input branch, has adjusted for various input variables. Table \ref{tab:hyper-network-architecture} shows the detailed architecture of hyper-networks. The columns ``operation''  represents the operation taken by this step, including convolutional operation, linear operation, layer normalization and etc. The column ``input'' refers to the source of input data for every operation. Among them, the items ``field data' and ``time'' represent the input meteorological field data and time information, respectively.  

\begin{table}[!htb]
  \centering
    \caption{Detailed Configuration of hyper-networks}
    \begin{tabular}{l|lcccc}
    \toprule
    &  name &operation & input & \textbf{$D_{in}$} & \textbf{$D_{out}$} \\
    \midrule
    \multirow{4}{*}{Embedding}
     & data\_em & conv1d & field data & -- & $M\times N$\\
     & time\em & position encoding & time & 1 & $1 \times N$\\
     & loc\_em & position encoding & data\_em + $W_L$ & $M\times N$ & $M\times N$ \\
     & input & add & data\_em+time\_em+loc\_em & $M\times N$ & $M\times N$ \\
     \midrule
     \multirow{7}{*}{n $\times$ MSA blocks} & multi\_att & 8 $\times$ self attention  & input & $M\times N$ & $M\times 8N$ \\
     & multi\_fc1 & linear & multi\_att & $M\times 8N$ & $M\times N$\\
     & norm1 & layer norm & multi\_fc1 + input & $M\times N$ & $M\times N$\\
      & fc1 & conv1d & norm1 &  $M\times N$ & $M\times N$  \\
     & act1 & gelu & multi\_fc1 & $M\times N$ & $M\times N$ \\
     & fc2 & conv1d & act1 &  $M\times N$ & $M\times N$  \\
     & norm2 & layer norm &  norm1 + fc2 & $M\times N$ & $M\times N$\\
     \midrule
     out block & out\_h & linear & norm2 & $M\times N$ & $M\times N$ \\
    \bottomrule
    \end{tabular}\label{tab:hyper-network-architecture}
\end{table}

\textbf{Physics networks}. 
The physics networks play a key role in bridging deep learning methods and atmospheric physics. It first utilizes the outputs of hyper-networks to calculate its parts of weights, which can incorporate spatial-temporal features extracted by hyper-networks. In addition, it is constructed based on MLPs, which accept coordinates and times as input and outputs corresponding values of variables. Physics networks that leverage the differentiable nature of neural networks to construct a partial differential equation (PDE) loss based on atmospheric dynamics and thermodynamic equations. This incorporation of physical constraints within the deep learning methods ensures that the forecasting results maintain physical interpretability. Table \ref{tab:physics-network-architecture} shows the detailed architecture of physics networks. For each output variable, corresponding physics networks with the same architecture are built. The items  'out\_h', 'time' and 'input coordinates' of column 'input' represent the outputs of hyper-networks, time information and input coordinates $x,y,t$.  

\begin{table*}[!htb]
  \centering
    \caption{Detailed Configuration of physics networks}
    \begin{tabular}{l|ccccc}
    \toprule
    name & operation & input & \textbf{$D_{in}$} & \textbf{$D_{out}$} & activation \\
    \midrule
    hyper\_map1 & linear & out\_h & $M\times N$ & $D_H+1 \times N$ & None\\
    hyper\_map2 & linear & out\_h & $M\times N$ & $D_H \times D_O+1$ & None\\
    coord\_fc1 &  hyper\_map1 & input coordinates & $B \times N$ & $B \times D_H$ & relu\\
    coord\_fc2 &  hyper\_map2 & coord\_fc1 & $B \times D_H$ & $B \times D_O$ & None\\
    time\_em & position encoding & time & $1$ & $1 \times N$ & None\\
    time\_fc & linear & time\_em & $1 \times N$ & $1 \times D_O$ & None\\
    ref\_em & position encoding & reference data & $1$ & $1 \times N$ & None\\
    ref\_fc & linear & ref\_em & $1 \times N$ & $1 \times D_O$ & None\\
    input & add & coord\_fc2 + time\_fc + ref\_fc & -- & $B \times D_O$ & None\\
    resmlp & residual MLP & input & $B \times D_O$ & $B \times D_O$ & relu\\
    out\_fc & linear & input + resmlp & $B \times D_O$ & $B \times 1$ & None\\
    out & add & reference data + out\_fc & $B \times 1$ & $B \times 1$ & None\\
    
    \bottomrule
    \end{tabular}\label{tab:physics-network-architecture}
\end{table*}

\end{appendices}

\end{onecolumn}

\end{document}